\documentstyle[12pt]{article}
\begin{document}
\begin{center}

{\bf The role of energy-momentum conservation \\
in emission of Cherenkov gluons}

\vspace{2mm}

I.M. Dremin\footnote{e-mail: dremin@lpi.ru}\\

{\it Lebedev Physical Institute, Moscow, Russia}

\end{center}

\vspace{1mm}

\begin{abstract}
The famous formula for the emission angle of Cherenkov radiation should be
modified when applied to hadronic reactions because of recoil effects. They 
impose the upper limit on the energy of the gluon emitted at a given angle. 
Also, it leads to essential corrections to the nuclear refractive index 
value as determined from the angular position of Cherenkov rings.
\end{abstract}

\noindent PACS: 24.85.+p; 25.75.Gz; 25.90.+k \\
Key words: Cherenkov radiation, gluons, partons, hadrons\\

\section{Introduction.}

The famous formula for the emission angle of Cherenkov radiation \cite{tf}
\begin{equation}
\cos \theta=\frac {1}{\beta n}   \label{cos}
\end{equation}
has been successfully used for Cherenkov photons. However, it was pointed out
already long ago \cite{gin} that energy-momentum conservation 
\begin{equation}
E=E_r+\omega, \;\;\; {\bf p}={\bf p_r}+{\bf k}
\end{equation}
asks for a
somewhat modified relation between the cone angle $\theta $ and the refractive
index of a medium $n$. Here, $E, {\bf p}$ and $E_r, {\bf p_r}$ are energies 
and momenta of the primary and recoiled emitters (electrons) while 
$\omega , {\bf k}$ correspond to the emitted quantum (photon).

Such a relation was first derived in \cite{gin}. The only new
element compared to usual energy-momentum conservation laws in vacuum is
the relation between the momentum of the emitted quantum $k$ and its energy
$\omega $ in a medium with the refractive index $n$: 
\begin{equation}
k=\omega n.                  \label{kon}
\end{equation}
Let us stress here that this relation is a consequence of the collective 
response of the medium to the quanta passing through it. This is 
described macroscopically by the refractive index $n$. Its quantitative 
microscopic interpretation is still missing. The whole process is 
considered in the rest system of the infinite medium.

The solution of energy-momentum conservation conditions for relativistic
($\beta \approx 1$)
electrons with this requirement leads to the following formula \cite{gin}:
\begin{equation}
\cos \theta = \frac {1}{n}\left (1+\frac {\omega }{2E}(n^2-1) \right ).\label{C}
\end{equation}
The second term in the brackets is positive for $n>1$
and diminishes the cone angle $\theta $. It is small for low energies
of photons $\omega \ll E$ and/or for the refractive index $n$ close to 1.
It would become important for energetic photons and large refractive indices.

In realistic situations with usual Cherenkov effect,
the correction term appears to be inessential for photons because
their energies are much less than the energy of electrons-emitters and the
refractive indices of most ordinary media are rather close to 1. Therefore
Eq. (\ref{cos}) can be used for Cherenkov photons with good accuracy. 
At the same time this correction happened to be important when considering 
the anomalous Doppler effect \cite{gfr}.

\section{Cherenkov gluons.}

The situation changes if one considers the emission of Cherenkov gluons
(first proposed in \cite{1, 2}) whose energies can be comparable by an
order of magnitude to the energy of a parton-emitter producing a jet.
Moreover, the partons-emitters are color charged and can be refracted in
the nuclear medium. We'll discuss this in Section 3, and first consider 
just the refraction of emitted gluons.

In contrast to Cherenkov photons propagating in ordinary media, the nuclear 
refractive index $n$ is unknown for Cherenkov gluons. Theoretical attempts
to estimate it are related either to its connection with the real parts of
hadronic forward scattering amplitudes using further assumption on 
applicability of their common features to gluons as well \cite{1, 2} or to
calculations of the polarization operator in simplified models \cite{kmwa}
(more discussion can be found in the review papers \cite{3, 4}).
They depend strongly on the unknown density of scattering centers in 
hadronic media. Therefore, it becomes the main goal 
to find $n$ by measuring the cone angle in high energy nucleus-nucleus 
collisions. From Eq. (\ref{C}) applied now to gluons the nuclear 
refractive index $n$ can be expressed as a function of the angle $\theta $:
\begin{equation}
n=\frac {E}{\omega }\left (\cos \theta - ((1-\omega /E)^2-\sin ^2\theta)^{1/2} 
\right ).                  \label{n}
\end{equation}
Considering real values of $n$ one gets from (\ref{n}) the upper
limit on the allowed range of gluon energies
\begin{equation}
\omega \leq E(1-\sin \theta ).
\end{equation}
This region becomes narrower at large cone angles. For gluons with larger 
energies $\omega $ the refractive index acquires the imaginary part. Therefore
the damping increases.

The relation (\ref{n}) is very simple for rather soft gluons satisfying 
the condition $\omega \ll 0.5E\cos ^2\theta $:
\begin{equation}
n \approx \frac {1}{\cos \theta }\left (1+\frac {\omega}{2E}\tan ^2\theta 
\right ).          \label{ln}
\end{equation}
One concludes that the traditional estimates of $n$ from the measured angle
$\theta $ using only the first term of (\ref{ln}) are lower than its actual 
value. The second term can be rather important
at large cone angles $\theta $ even at small gluon energies $\omega \ll E$.

Let us relate these formulas to experimental findings. The cone structure
of hadron emission around the away-side jets has been observed in several
experiments on nucleus-nucleus collisions at RHIC 
\cite{5a, 6a, adar, jia, ajit, pr}. The cone angle values are quite large
so that the second term in Eq. (\ref{ln}) is important. They vary between
60$^o$ and 70$^o$. This undefiniteness is related to different methods
of rings presentation and to their finite widths. Surely, this range will
become shorter with newly coming data. We present our estimates for two
values of $\cos \theta $ equal to 0.5 (60$^o$) and 0.342 (70$^o$) to
demonstrate how important are the corrections obtained above.

At $\theta = 60^o$ one gets $\omega < 0.134 E, \;\; n=2(1+1.5\omega /E) < 2.4$.
If the total energy of the away-side jet is $E=5$ GeV, it implies
$\omega < 0.67$ GeV, i.e. only states in the low-mass wing of $\rho $-resonance
can be created.

At $\theta = 70^o$ one gets $\omega < 0.06 E, \;\; n=2.9(1+3.8\omega /E) < 3.6$.
For $E=5$ GeV we get $\omega < 0.3$  GeV, i.e. in this case only very low energy 
tail of $\rho $-resonance can play a role so that one would expect low intensity
for such a process. The value $n=3$ used in \cite{3, 4} lies in between
the above limits.

\section{Discussion.}

In electrodynamics, the charged current (electrons) differs strongly from
neutral emitted radiation (photons). Therefore, the medium impact on these 
components in Cherenkov effect is different. The refractive index is introduced 
for photons while the electron motion is considered as almost undisturbed. 
It is supported by smallness of energy losses for Cherenkov radiation.
That is why the relation (\ref{kon}) was used for photons only.  
In general, the refractive index depends on energy but it is constant with 
high precision for visible light.

In chromodynamics, the partons-emitters (quarks or gluons) are colored as well 
as the emitted gluons. They differ only by their energies. In principle, all
three of them can be refracted by the nuclear medium. If there is no 
dispersion (energy dependence) of the nuclear refractive index, the nuclear
Cherenkov effect is impossible. In above treatment,
it was implicitly assumed that the nuclear refractive index is constant and
exceeds 1 for comparatively soft emitted gluons while for emitters with high
energies it is inessential, i.e. close to 1. Then Eq. (\ref{C}) is valid.
This assumption can be justified 
only relying on possible analogy of partonic properties to hadronic reactions 
used earlier (see reviews in \cite{3, 4}).

It is well known for all studied hadron collisions that the real part of the 
forward scattering amplitude depends on energy. It is positive in the low-mass 
wings of the Breit-Wigner resonance regions, then it is negative at higher 
energies and becomes again positive at extremely high energies but the latest 
region is not considered right now. Since excess of $n$ over 1 is proporional to 
this real part, one can argue \cite{3, 4} that Cherenkov gluons can excite 
the collective modes of the nuclear medium within the resonance region. Thus 
the energy behaviour of the nuclear refractive index $n$ plays an important 
role for such emission\footnote{Its energy dependence can be easily taken into 
account as it was done, e.g., in \cite{dne} but above we used constant $n$
treating it as an effective average value in the resonance region.}. Since 
energies of partons-emitters are higher than those of Cherenkov gluon, one can 
neglect their refraction because $n$ is close to 1 (and below it) in this 
region above resonances. Then the formula (\ref{C}) is justified for Cherenkov 
gluons if applied to results of trigger experiments at RHIC.

What concerns the non-trigger experiments with extremely high energy partons 
at LHC (in the third region described above), the refractive indices of all
three participants can be taken into account. It leads to the simple
generalization of above formulas so that Eq. (\ref{n}) is replaced by
\begin{equation}
n=\frac {E}{\omega }n_0\left (\cos \theta - (
\frac {n_r^2}{n_0^2}(1-\omega /E)^2 - \sin^2\theta)^{1/2} \right ),      \label{nn}
\end{equation}
where $n_0$ and $n_r$ are the refractive indices of the initial and recoiled
partons, correspondingly.

In particular, if the energies of the initial and recoiled partons
are approximately the same then their refractive indices are almost equal
($n_0 \approx n_r$), and one should use Eq. (\ref{ln})
with $n$ replaced by the ratio $n/n_0$. At high energies the refractive 
indices decrease with energy increasing (see \cite{3, 4}). Thus $n_0$ is much
closer to 1 than the refractive index $n$ of the gluon emitted with lower 
(but still very high) energy. The above formulas work quite well in this 
region also.

To conclude, we have shown that energy-momentum conservation laws play a
crucial role in determining main characteristics of emission of Cherenkov
gluons. They should be carefully taken into account in analysis and 
interpretation of experimental data about the ring-like structure recently
observed in high energy nucleus-nucleus collisions. Such analysis must
reveal the values of the nuclear refractive index and their energy
dependence. Herefrom one would conclude about the state of the matter
formed in high energy nucleus-nucleus collisions.

\bigskip 

{\bf Acknowledgements.}

This work is supported in part by the RFBR grants 06-02-17051-a, 06-02-16864-a.

\end{document}